\documentclass[iop]{emulateapj}
\usepackage{graphics}
\usepackage{rotating}

\newcommand{\ltsimeq}{\la}
\newcommand{\gtsimeq}{\ga}

\newcommand{\msun}{M$_{\odot}$}

\newcommand{\HI}{H{\sc i}}

\shortauthors{McQuinn et al.}
\shorttitle{Observational Constraints on the Molecular Gas Content in Nearby Starburst Dwarf Galaxies}

\begin{document}
\title{Observational Constraints on the Molecular Gas Content in Nearby Starburst Dwarf Galaxies\footnote{Based on observations made with the NASA/ESA Hubble Space Telescope, obtained from the Data Archive at the Space Telescope Science Institute, which is operated by the Association of Universities for Research in Astronomy, Inc., under NASA contract NAS 5-26555.}}
\author{Kristen.~B.~W. McQuinn\altaffilmark{1}, 
Evan D. Skillman\altaffilmark{1},
Julianne J.~Dalcanton\altaffilmark{2},
Andrew E.~Dolphin\altaffilmark{3},
John M.~Cannon\altaffilmark{4},
Jon Holtzman\altaffilmark{5},
Daniel R.~Weisz\altaffilmark{2},
Benjamin F.~Williams\altaffilmark{2}
}

\altaffiltext{1}{Department of Astronomy, School of Physics and
Astronomy, 116 Church Street, S.E., University of Minnesota,
Minneapolis, MN 55455, \ {\it kmcquinn@astro.umn.edu}} 
\altaffiltext{2}{Department of Astronomy, Box 351580, University 
of Washington, Seattle, WA 98195}
\altaffiltext{3}{Raytheon Company, 1151 E. Hermans Road, Tucson, AZ, 85756}
\altaffiltext{4}{Department of Physics and Astronomy, 
Macalester College, 1600 Grand Avenue, Saint Paul, MN 55105}
\altaffiltext{5}{Department of Astronomy, New Mexico State University, Box 30001-Department 4500, 1320 Frenger Street, Las Cruces, NM 88003}

\begin{abstract}
Using star formation histories derived from optically resolved stellar populations 
in nineteen nearby starburst dwarf galaxies observed with the  
\textit{Hubble Space Telescope}, 
we measure the stellar mass surface densities of stars newly formed in the bursts.
By assuming a star formation efficiency (SFE), we then calculate the 
inferred gas surface densities present at the onset of the starbursts.
Assuming a SFE of 1\%, as is often assumed in normal star-forming galaxies, and
assuming that the gas was purely atomic, translates to very high \HI\ surface
densities ($\sim$ $10^2-10^3$ \msun~pc$^{-2}$), which are much higher than
have been observed in dwarf galaxies. This implies either higher values of SFE
in these dwarf starburst galaxies or the presence of significant amounts of 
H$_2$ in dwarfs (or both). Raising the assumed SFEs to 10\% or greater (in line with observations of more massive starbursts associated with merging galaxies), still results in \HI\ surface densities higher than observed in 10 galaxies. Thus, these observations appear to require that a significant fraction of the gas in these dwarf starbursts galaxies was in the molecular form at the onset of the bursts. Our results imply molecular gas column densities in the range $10^{19}-10^{21}$ cm$^{-2}$ for the sample.
In those galaxies where CO observations have been made, these densities correspond to values of the CO$-$H$_2$ conversion factor (X$\rm_{CO}$) in the range $>3-80\times10^{20}$ cm$^{-2}$ (K km s$^{-1}$)$^{-1}$, or up to $40\times$ greater than Galactic X$\rm_{CO}$ values.
\end{abstract} 


\section{Introduction\label{intro}}
Star formation rates (SFRs) are known to correlate with the surface density of cold gas. This correlation, commonly referred to as the Kennicutt-Schmidt law, was seen using star formation (SF) and gas tracers integrated on galaxy scales \citep{Schmidt1959, Kennicutt1989, Kennicutt1998b}. The direct correlation of SF and molecular gas has been seen on sub-galactic spatial scales as well, using maps of atomic and molecular gas \citep[e.g.,][]{Martin2001, Wong2002, Bigiel2008, Rahman2012}. However, the variables and conditions that govern when and how much gas is converted into stars remains unclear \citep[e.g.,][and references therein]{Tan2000, Martin2001, Wong2002, Leroy2008}. In addition, the role atomic \HI\ gas plays in driving the SFR is also poorly understood, particularly in dwarf and low surface brightness galaxies which lack significant CO detections that trace molecular gas \citep[e.g.,][and references therein]{Taylor1998, ONeil2003, Leroy2005}.

While most studies approach this analysis based on the current SFRs and gas surface densities, studies of optically resolved stellar populations can provide a longer temporal baseline by probing the SF as a function of time (i.e., SFR(t)). Using \textit{Hubble Space Telescope} ($HST$) observations, \citet{McQuinn2010a,McQuinn2010b} reconstructed the star formation histories (SFHs) of 19 nearby, starburst dwarf galaxies. These temporally resolved SFHs enable a calculation of the total stellar mass of stars newly formed in these systems over the duration of the burst events (i.e., a few 100 Myr). As the stellar mass was formed from the gas reservoirs of the galaxies, this calculation provides a way to estimate the gas surface densities present when the star forming episode began.

One way to quantify the amount of gas that is converted into stars is through the star formation efficiency (SFE) parameter, which estimates the fraction of gas that is converted into stars in a given area during a fiducial time period. 
The SFE defined in this way is dimensionless and based on the direct connection between gas and the stars formed from the gas. Models of SF, based on different SF laws and SF thresholds, assume different 
SFEs in their calculations, or conversely, predict SFEs given a set of 
inputs \citep[e.g.,][]{Tan2000, Parmentier2009, Cote2011}. 
Measurements of the SFE are problematic, as the measurements of the gas 
being converted into stars and of the stars produced from that gas cannot
be made simultaneously.  In practice, observations of SFEs usually assume
typical observed gas surface densities or gas surface densities in the 
vicinity of recent star formation. 

SFE's have been studied in a variety of galaxy types and environments including, in order of increasing SFRs, dwarf galaxies \citep{Bigiel2008}, normal disk galaxies \citep{Kennicutt1998b, Bigiel2008, Rahman2012}, luminous infrared galaxies \citep{Young1986, Solomon1988, Sanders1991}, and massive starburst galaxies \citep{Kennicutt1998b}. The SFE values range from $\sim1$\% to values approaching $-\sim100$\% in massive starbursts \citep{Kennicutt1998b}, with the majority of normal star forming galaxies (i.e., dwarfs and spirals) having lower SFEs in the $\sim1-10$\% range. SFEs higher than 10\% have been reported in more massive starburst galaxies \citep[e.g.,][]{Young1986, Solomon1988, Kennicutt1998b, Sanders1991, Bigiel2008}, possibly as a result of higher gas densities according to the Kennicutt-Schmidt Law. However, on small physical scales SFEs have also been argued to be roughly invariant in any supersonically turbulent medium \citep{Krumholz2005}. 

In this study, we use SFHs derived from resolved stellar 
populations in 19 nearby, \HI\ dominated, starburst dwarf galaxies 
\citep{McQuinn2010a, McQuinn2010b} in order to investigate the gas content 
in these galaxies prior to their bursts. For the derived stellar mass surface densities of the stars formed in the bursts, we assume different SFEs to infer the gas surface densities present at the onset of the starbursts. Further, we investigate the implications these gas surface densities have for the amount molecular gas that may have been present in these systems. The paper is organized as follows. \S2 summarizes the observations, data processing, and SFH reconstruction technique. \S3 calculates the surface density both of the SFR and of the stellar mass. \S4 applies different SFEs to infer the gas surface densities present when the starbursts began and \S5 estimates CO$-$H$_2$ conversion factors. \S6 summarizes our conclusions.

\section{The Galaxy Sample, Stellar Populations, and SFHs\label{obs}}
\begin{figure}[h]
\plotone{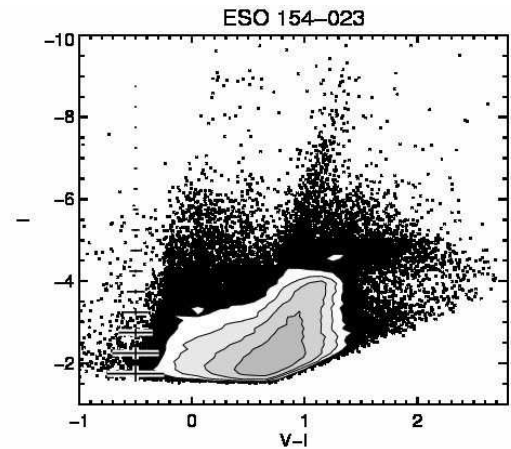}
\caption{The CMD of ESO~154$-$023 showing the typical photometric quality of the data used in the analysis. Photometric uncertainties are shown in the left hand side of the CMD. Contours are used to show the photometric structure of the stellar populations in the red giant branch and lower main sequence at levels of 75, 150, 300, and 600 point sources per magnitude-color bin. The blue and red helium burning sequences are unambiguous signs of recent SF. The CMDs for the entire sample can be found in \citet{McQuinn2010a}. }
\label{fig:cmd}
\end{figure}

In Table~\ref{tab:galaxies} we summarize various physical properties of the 19 nearby starburst dwarf galaxies that comprise our sample. This sample was previously studied by \citet[][and references therein]{McQuinn2010a, McQuinn2010b} using archival $HST$ V and I band observations from either the Advanced Camera for Surveys or Wide Field Planetary Camera 2 instruments. The galaxies all lie close enough for the $HST$ observations to resolve the stellar populations into individual stars (i.e., D $\ltsimeq 5$ Mpc). A full description of the observations and data processing can be found in \citet{McQuinn2010a}. For this paper we use the \citet{McQuinn2010a} photometry, which was measured by DOLPHOT or HSTphot \citep{Dolphin2000} from the $HST$ pipeline processed and cleaned images. In Figure~\ref{fig:cmd}, we present an example of the photometric quality in the color magnitude diagram (CMD) for ESO~154$-$023, whose data reach our minimum required photometric depth. Representative uncertainties are shown per magnitude by crosses in the left hand side of the CMD. CMDs can be found in \citet[][Figure~2]{McQuinn2010a} for the majority of the sample and in \citet{Weisz2008} for two galaxies (DDO~165 and Ho~II). 

The SFHs (i.e., the rate of star formation across time $SFR(t,Z)$) were reconstructed by \citet{McQuinn2010a} by fitting the observed CMDs \citep{Dolphin2002}. This well established technique uses stellar evolutionary models \citep{Marigo2007} to model the observed CMD as a linear combination of different age and metallicity stellar populations for an assumed initial mass function (IMF). As described in \citet{McQuinn2010a}, the SFHs were reconstructed assuming a binary fraction of 0.35 and a Salpeter IMF \citep{Salpeter1955}, and allowing the models to fit for distance and extinction. The metallicity was constrained to increase with time, except in cases where the photometry reached a full magnitude below the red clump. For a full discussion of the SFHs, the time binning used, and uncertainties we refer the reader to \citet{McQuinn2010a}. We scaled down the SFRs by a factor of 1.59 \citep{Bruzual2003} to transform the results to a Kroupa IMF \citep{Kroupa2001}, thus enabling a direct comparison with other studies in the literature on SFR and SFR densities \citep[e.g.,][]{Bigiel2008}.

\begin{figure}[h]
\plotone{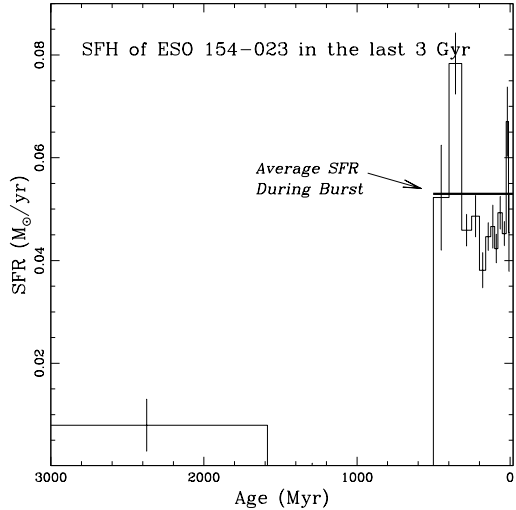}
\caption{The SFH of ESO~154$-$023 during the last 3 Gyr. The SFR is calculated assuming a Kroupa IMF. The SFR is higher over the past 500 Mr, preceded by a period ($\sim2.5$ Gyr) of quiescence when the SFR was a factor of $\sim5$ lower. This pattern of sustained elevated SFR at recent times is seen across the entire starburst sample. In this study, we measure the $\Sigma\rm_{SFR}$ and calculate the stellar mass surface density during this period of elevated SF. The bin size along the horizontal axis was chosen based on temporal resolution achieved from SFHs reconstructed from synthetic CMDs of similar completeness. The same binning is used for each galaxy in the sample. The error bars include both systematic uncertainties and statistical uncertainties estimated from Monte Carlo simulations.  The uncertainties are typical of the SFH of the sample.}
\label{fig:sfh}
\end{figure}

In Figure~\ref{fig:sfh}, we present an example of the resulting SFH of ESO~154$-$023 over the last 3 Gyr. For this starburst system, the recent SFRs are elevated over the average value during the most recent $\sim450 \pm50$ Myr \citep{McQuinn2010b}. This pattern of sustained elevated SFRs at recent times is seen across the starburst sample. For the current analysis, we focus on measuring the SFR surface density over the duration of the starbursts, and on calculating the stellar mass surface density for the stars formed during the same time period.

\section{Stellar Mass Surface Densities of Newly Formed Stars \label{sfr_area}}

SF intensity is frequently parameterized by calculating the SFR surface density, $\Sigma\rm_{SFR}~\equiv$ SFR area$^{-1}$ (\msun~yr$^{-1}$~kpc $^{-2}$) \citep[e.g.,][]{Meurer1997,Lanzetta2002, Kennicutt2005}. This calculation of $\Sigma\rm_{SFR}$ depends upon the timescale over which one measures the SFRs. For example, the SFR calculated from H$\alpha$ flux \citep[e.g.,][]{Kennicutt1998a, James2004} is the SFR averaged over the past few Myr, whereas the SFR calculated from UV emission \citep{Meurer1997} is an average over the past few hundred Myr. If the SFR remains constant over the two time periods, the averages are equivalent, but variations in the SFR over time can make the different values difficult to compare. The optically resolved stellar populations and temporally resolved SFHs allow us to select the timescale over which to calculate $\Sigma\rm_{SFR}$. We chose to calculate $\Sigma\rm_{SFR}$ over the duration of a burst (see Figure~\ref{fig:sfh}). The durations and SFRs averaged over these timescales, listed in Table~\ref{tab:galaxies}, were compiled from \citet{McQuinn2010b}.

There are a number of possible choices for normalizing the SFR to a fiducial area. In nearby systems, both \citet{James2004} and \citet{Kennicutt2005} use the area of the currently active star forming regions as traced by H$\alpha$ emission. The $\Sigma\rm_{SFR}$ calculated in this way follows a Gaussian distribution for nearby galaxies ranging from $10^{-4}$ to $10^2$ \msun~yr$^{-1}$~kpc$^{-2}$, with massive starburst galaxies generally falling in an extended tail of the distribution above $10^{-1}$ \msun~yr$^{-1}$~kpc$^{-2}$. Alternatively, the area has been defined as the half-light radius of a galaxy in the UV \citep[i.e., radius $\sim 0.1-10$ kpc;][]{Meurer1997}, or as the optical disk of a galaxy measured to the B-band 25 mag arcsec$^{-2}$ isophotal radius \citep[i.e., radius $\sim$ kpc scales;][]{Schmidt1959, Kennicutt1998b}. In a comparative study using \HI, UV, and optical data, \citet{Wyder2009} use the minimum area for which all wavelengths showed a detection and the $\Sigma\rm_{SFR}$ was calculated using an average SFR based on UV surface brightness over the adopted area. 

Over the duration of a starburst, in many dwarf galaxies, SF has been shown to be widely distributed across the disks of the galaxies, rather than solely concentrated at the very center (McQuinn et al.\ in prep.). Therefore, for this study, we use the optical extent of the galaxies included in the $HST$ observational fields of view. The $\Sigma\rm_{SFR}$ calculated from this area represents a measurement of the SF activity in the optical disk over the life of a starburst. These values could potentially be compared with SFR surface densities calculated for high redshift bursting dwarf galaxies, should observations of this type become available. Note that for our analysis of $\Sigma\rm_{SFR}$, this choice in area represents a conservative assumption. In addition, if the gas is assumed to be clumped (as expected), then this will result in even higher values of precursor gas surface density.

The optical areas of the galaxies were calculated based on the major diameter measured from the extent of the B-band 25 mag arcsec$^{-2}$ isophotal radius and the ellipticity of the galaxy \citep[e.g., D$_{25}$;][see Table~1]{Karachentsev2004}. These D$_{25}$ measurements were corrected for inclination angle and Galactic extinction following the prescription by \citet{deVaucouleurs1976}. Included in Table~\ref{tab:galaxies} are the deprojected D$_{25}$ areas covered by the $HST$ fields of view and the equivalent areas in kpc$^2$ for the sample, accounting for distance to the system. In eleven galaxies, the observational field of view covered the extent of the galaxy out to the D$_{25}$ limits. In the remaining cases, the angular area of the observations did not extend to this surface brightness limit. For these galaxies, we normalized the SFR by the area covered by the $HST$ observations corrected for inclination. We discuss the impact of these smaller fields of view on our results below.

\begin{figure}[h]
\plotone{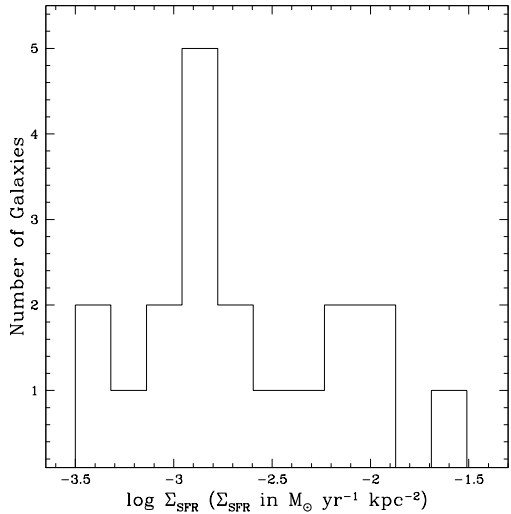}
\caption{The histogram shows the range of $\Sigma\rm_{SFR}$ found in the sample. The values are lower than typical starburst SFR surface densities of more massive, H$_2$ rich galaxies found in the literature \citep[e.g.,][]{Kennicutt1998b, Gao2004}, both because the sample consists of lower mass galaxies that are \HI\ dominated and because we use the larger area of the optical disk as our baseline area. These $\Sigma\rm_{SFR}$ are similar to measured values for spiral disks galaxies when averaged over sub-kpc scales \citep{Bigiel2008}. Thus, while the sample is comprised of dwarf galaxies, the bursting SF makes their SFR surface densities comparable to the higher end of more massive spiral disk galaxies.}
\label{fig:histo_sfr}
\end{figure}

In Figure~\ref{fig:histo_sfr}, we present a histogram of our measured $\Sigma\rm_{SFR}$ ranging from  $10^{-3.5}$ to $10^{-1.5}$ \msun~yr$^{-1}$~kpc$^{-2}$. Values for individual galaxies are listed in Table~\ref{tab:densities}. The range of values in this starburst sample is lower than the 0.1 \msun~yr$^{-1}$~kpc$^{-2}$ threshold set for starbursts in more massive galaxies based on H$\alpha$ derived SFRs \citep{Kennicutt2005}. $\Sigma\rm_{SFR}$ is generally higher in more massive disk galaxies \citep{Kennicutt1998b, Bigiel2008}. However, $\Sigma\rm_{SFR}$ values from different studies are difficult to compare as the adopted areas and the methods of measuring the SFRs often vary. 

The total stellar mass formed during the bursts can be calculated directly from the SFHs by integrating the SFRs(t) over the duration of the bursts. Applying the same de-projected physical areas in Table~\ref{tab:galaxies}, we calculate the stellar mass surface density of stars formed during the burst ($\Sigma\rm_{*,burst}$), ranging from 0.2 to 13 \msun~pc$^{-2}$. Values for individual galaxies are listed in Table~\ref{tab:densities}. For the galaxies where the observational field of view did not reach the isophotal radius of 25 mag arcsec$^{-2}$, our analysis applies to only the central regions of the systems. The difference between this central value of the stellar mass surface density of the newly formed stars and the value for the entire optical disk is difficult to estimate as the SFRs in the outer regions of galaxies do not scale linearly with area. This may be an important difference in the three galaxies where the $HST$ field of view covers less than half of the optical disk (see Table~\ref{tab:galaxies}), but will not impact our results significantly in the remaining systems. Note also that the stellar mass values are dependent on the form of the IMF assumed in deriving the SFHs. If we adopted a Salpeter IMF \citep{Salpeter1955}, these stellar mass values would be larger by a factor of 1.59.

\section{SFEs and Inferred Initial Gas Surface Densities \label{gas_area}}
Having measured $\Sigma\rm_{*,burst}$, we can infer the hydrogen gas surface density (hereafter $\Sigma\rm_{gas}$) present at the onset of the starbursts, for an assumed SFE and an assumed primordial Helium abundance of 1.33 \citep{Cyburt2008, Komatsu2011}. This simple model assumes that the initial hydrogen gas mass is a multiple of the stellar mass formed in the burst in the stars (i.e., $\Sigma\rm_{gas} \equiv (\Sigma\rm_{*,burst}/1.33)/$SFE). The SFEs in dwarf galaxies are typically between $1-3$\%, comparable to the SF in outer disks of spiral galaxies. However, while the starburst dwarf galaxies in our sample have low masses and low metallicities typical of dwarf galaxies, their $\Sigma\rm_{SFR}$ are comparable to or higher than the values of $\Sigma\rm_{SFR}$ for typical spiral galaxies which show SFEs between $5-10$\% \citep{Bigiel2008, Rahman2012}. 

Based on the previous results for both dwarf and spiral galaxies, we might expect SFEs for our sample to lie in the range of  $1-10$\%. Starting with the lower limit SFE value of $1\%$, typical of the lower SFRs typically found in the outer regions of spirals and in dwarf galaxies \citep{Kennicutt1998b, Bigiel2008}, the inferred $\Sigma\rm_{gas}$ ranges from $25 - 900$ \msun\ pc$^{-2}$. However, if we assume that the ISMs in these dwarf galaxies are \HI\ dominated, as suggested by observations \citep[e.g.,][]{Leroy2008}, this derived range of gas surface densities includes densities far greater than any \HI\ gas surface densities observed to date. The typical value of the maximum $\Sigma\rm_{HI}$ observed in dwarf galaxies is limited to the range of $\sim5-10$ \msun\ pc$^{-2}$ \citep{Swaters2002} with a sharp saturation at $\sim10$ \msun\ pc$^{-2}$ \citep{Wong2002, Bigiel2008} when measured on sub-kpc scales. These numbers will be lower if averaged over larger physical scales. The implication is that the SFE must be higher than 1\%, or that not all of the hydrogen is in the atomic phase (or both). Because the highest stellar mass surface densities of newly formed stars derived assuming a SFE of 1\% are roughly 100 times larger than the highest observed HI surface mass densities, solving this problem would require SFEs of order 100\%, which are very unlikely. 

Atomic hydrogen gas surface densities higher than any observed would be possible if the triggering event of the starburst provided additional gas to the systems \citep{Ekta2010}, temporarily increasing $\Sigma\rm_{gas}$ far above what is typically observed in \HI. As none of the galaxies in the sample show evidence of an obvious merger, gas infall would be the most likely source of the additional mass. However, \HI\ maps, available for roughly half of the sample \citep{Begum2008, Bigiel2008, Warren2012}, do not show highly disturbed \HI\ morphologies, as would be expected from a gas infall event. We will go forward assuming a higher SFE of 10\% and see what that implies about the gas surface densities and the molecular gas fractions.

\begin{figure}[h]
\plotone{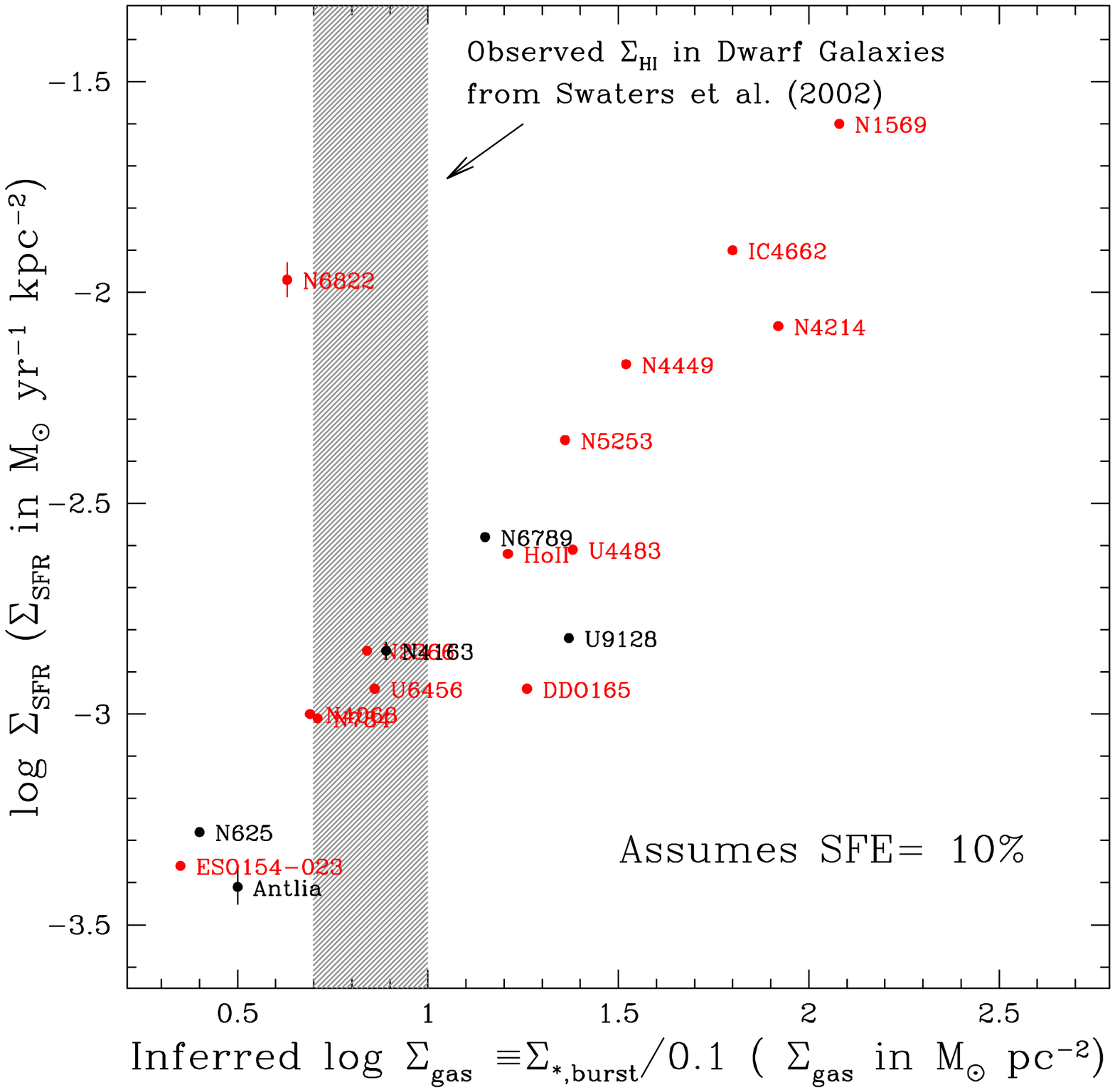}
\caption{Assuming a SFE of 10\%, the inferred hydrogen gas surface densities present at the onset of the bursts range from $2-90$ \msun~pc$^{-2}$, after correcting for the abundance of Helium. The galaxies plotted in red have ongoing bursts, while those plotted in black are fossil bursts. Uncertainties in the $\Sigma\rm_{SFR}$ are smaller than the size of the points, except where drawn in the Figure. We adopted one value of SFE, but the SFE is likely to be lower for the systems with log $\Sigma\rm_{SFR} <-3$, based on measurements in galaxies with similar SFR surface densities \citep{Bigiel2008}. 
Therefore, the gas surface densities in these systems are likely to be higher. 
For the galaxies with log $\Sigma\rm_{SFR} \gtsimeq -2.9$, even with assuming a SFE of 10\%, 
the inferred $\Sigma\rm_{gas}$ is higher than the observed range in 
maximum values of $\Sigma\rm_{HI} = 5-10$ \msun\ pc$^{-2}$ observed in dwarf galaxies \citep{Swaters2002}, with a sharp saturation at $\sim9$ \msun\ pc$^{-2}$ \citep{Bigiel2008}). Thus, we conclude that a significant fraction of the gas is in molecular form and 
remains unobserved in most of these galaxies. Further, we estimate the X$\rm_{CO}$ factor to range from 3 to 80 $\times 10^{20}$ cm$^{-2}$ (K km s$^{-2}$)$^{-1}$ based on the inferred molecular gas surface densities and CO observations, where available, in the literature.}
\label{fig:surface_densities}
\vspace{12pt}
\end{figure}

Figure~\ref{fig:surface_densities} shows the relationship between the observed SFR surface density and the inferred gas surface density with an assumed SFE of $10\%$. The points deviate from a straight line because systems with similar average SFR densities can have different stellar mass surface densities of recently formed stars depending on the duration of each burst. While 5 galaxies host ``fossil'' bursts, the remaining 14 host on-going starburst activity. Thus, many of the gas surface density values will increase as the bursts mature. Figure~\ref{fig:surface_densities} demonstrates that even with a higher SFE of 10\%, half the sample have inferred $\Sigma\rm_{gas}$ that exceed any observed atomic hydrogen gas surface densities to date \citep[$\Sigma\rm_{HI} \sim10$ \msun\ pc$^{-2}$;][]{Swaters2002, Wong2002, Bigiel2008}. Conversely, if we assume that $\Sigma\rm_{gas}$ is $\ltsimeq10$ \msun\ pc$^{-2}$ based on \HI\ observations, we infer a minimum SFE of 3\% and a maximum SFE far exceeding 100\%. Thus, it appears that the calculated stellar mass surface densities of newly formed stars imply that the typical dwarf galaxy has a significant reservoir of molecular hydrogen.  

Assuming the initial gas mass is some combination of atomic and molecular gas, the values of the inferred $\Sigma\rm_{gas}$ seen in Figure~\ref{fig:surface_densities} suggest that a significant fraction of the gas is in molecular form in 10 of the galaxies. If we adopt both a SFE of 10\% and a $\Sigma\rm_{HI} = 10$ \msun\ pc$^{-2}$, then the inferred molecular gas surface density ($\Sigma\rm_{H_2}$) ranges from $\sim1-80$ \msun\ pc$^{-2}$ or an equivalent average H$_2$ column density of $10^{19}-10^{21}$ cm$^{-2}$; values for the individual galaxies are listed in Table~\ref{tab:densities}. This range would be higher if a lower SFE were assumed (this is likely the case for the systems with log $\Sigma\rm_{SFR} <-3$). Our estimated values of $\Sigma\rm_{H_2}$ are averages for the optical disk; regions of even higher molecular gas densities would be expected in star-forming regions.

Our inferred $\Sigma\rm_{H_2}$ can be compared with known measurements of $\Sigma\rm_{H_2}$ in more massive disk galaxies with similar $\Sigma\rm_{SFR}$. For example, $\Sigma\rm_{H_2}$ has been measured to range from $\sim3-70$ \msun~pc$^{-2}$ in the inner regions of lower metallicity spiral galaxies \citep{Bigiel2008, Leroy2011}. Somewhat higher H$_2$ column density estimates of up to $\sim10^{22}$ cm$^{-2}$ (i.e., 100 \msun\ pc$^{-2}$) have been found in the SMC from infrared dust measurements on 200 pc scales \citep{Bolatto2011}. In the central regions of 14 normal spiral galaxies, \citet{Rahman2012} have measured $\Sigma\rm_{H2}$ to range from $20-1000$ \msun\ pc$^{-2}$, however giant molecular clouds in the Milky Way have a mean hydrogen molecular gas surface density of 42 \msun\ pc$^-2$, with variations up to $80-120$ \msun\ pc$^{-2}$ \citep{Heyer2009}. Based on these comparisons, the estimated values of $\Sigma\rm_{H_2}$ (ranging from $\sim1-80$ \msun\ pc$^{-2}$) and the corresponding fraction of molecular to total gas are in agreement with molecular gas surface densities and with molecular to atomic ratios measured in galaxies with similar $\Sigma\rm_{SFR}$. In addition, the typical timescale to deplete the molecular gas by star formation in a disk galaxy is $\sim2$ Gyr \citep{Bolatto2011, Rahman2012}. This depletion timescale is up to $4\times$ longer than the duration of the starburst events indicating that typical disk galaxies will contain sufficient reservoirs of molecular gas over the timescales of a starburst to fuel and sustain the elevated levels of SF. 

\section{Limits on the CO-H$_2$ Conversion Ratio\label{CO_conversion}}
While the atomic \HI\ component of galaxies is easily measured using 21 cm observations, the molecular H$_2$ component is generally measured indirectly by measuring emission from the CO molecule and then inferring a ratio of the CO to H$_2$ molecules (i.e., the CO-H$_2$ conversion factor, X$\rm_{CO}$). Estimating H$_{2}$ masses from CO emission has proven more difficult in dwarf galaxies for two reasons. First, there are well known difficulties in detecting molecular gas in low metallicity environments \citep[e.g.,][and references therein]{Bolatto2008}. Since few dwarf galaxies have CO emission detections, conversions from CO emission to H$_2$ masses are often only upper limits. Second, X$\rm_{CO}$ is uncertain for low metallicity systems. Historically, X$\rm_{CO}$ was based on measurements at higher metallicities and application to lower metallicity systems was uncertain. Indeed, the conversion factor has been measured to be a strong function of metallicity below 12$+$log(O/H) $\sim8.4-8.2$ \citep{Leroy2011}, the upper end of the metallicity range of our sample. Thus, even in cases with CO emission detections, conversion to an H$_2$ mass is uncertain for low metallicity dwarfs.

We can constrain the values of X$\rm_{CO}$ for a few systems in our sample by combining our inferred values of $\Sigma\rm_{H2}$ with confirmed CO detections (or limits on the CO emission) from the literature. While most galaxies in our sample have very low metallicities (i.e., 12$+$log (O/H)$\ltsimeq8.0$; see Table~\ref{tab:galaxies} for individual values) where CO is completely non-detected \citep[e.g.,][]{Taylor1998}, a few of the more luminous galaxies in our sample with somewhat higher metallicities (i.e., 12$+$ log(O/H) $\sim8.1-8.4$) do have CO detections (e.g., NGC~1569, \citet{Young1984}; NGC~4449 and NGC~4214, \citet{Tacconi1985}; NGC~5253, \citet{Turner1997}; NGC~6822, \citet{Wilson1992}). In Table~\ref{tab:densities}, we list confirmed CO detections from the literature for five galaxies and CO upper limits for an additional three galaxies. Using these CO observations and our estimates of N$\rm_{H_2}$ in the galaxies with inferred molecular gas content, we calculate X$\rm_{CO}$ to range from ($>3 - 80$)$ \times10^{20}$ cm$^{-2}$/(K km s$^{-1}$). The lower end of this range is $\sim1.5\times$ greater than Galactic measurements while the upper end is $40\times$ greater. These estimated values of X$\rm_{CO}$ are in agreement with measured values of $\sim20\times10^{20}$ and $\sim45\times10^{20}$  cm$^{-2}$/(K km s$^{-1}$) found for two dwarf galaxies in the Local Group from infrared dust emission \citep{Leroy2011} and the lower limit of $10\times$ Galactic X$\rm_{CO}$ values placed on a low metallicity (12$+$[O/H]$\sim7.67$) Local Group dwarf galaxy DDO~154 \citep{Komugi2011}.

Since nearby, metal poor, dwarf galaxies are often considered the prototypes of more massive galaxies at high redshift, it is also interesting to compare our estimated X$\rm_{CO}$ values to measurements in low metallicity systems at high redshift. In a sample of star forming galaxies in the mass range $10^{10}-10^{11}$ \msun, one to two orders of magnitude more massive than any of the galaxies in our sample, \citet{Genzel2011} found the CO luminosity to molecular gas mass conversion factors to be $\sim2.5-14 \times$ Galactic values in a metallicity range of 12$+$[O/H] $\sim8.1-8.4$, with higher values expected in lower mass galaxies. For comparison, for the lower mass galaxies in our sample which span the range in metallicities from $12+$[O/H]$\sim7.4-8.4$ (see Table~\ref{tab:galaxies}), our estimated X$\rm_{CO}$ factors include values up to $40\times$ greater than Galactic X$\rm_{CO}$ values. 

\section{Conclusions \label{conclusions} }
We have used optical imaging of resolved stellar populations obtained from the $HST$ data archive to measure the SFRs and stellar mass surface densities of stars formed during starburst events in a sample of 19 nearby starburst dwarf galaxies. The SFR surface densities of these dwarf galaxies are comparable to the higher end of SFR surface densities of more massive spiral galaxies while still having low metallicities typical of dwarf systems.   

By assuming different SFEs, the stellar mass surface densities were used to infer gas surface densities present at the onset of the bursts. Using a SFE of 1\% and assuming $\Sigma\rm_{gas} \sim \Sigma\rm_{HI}$, all inferred gas surface densities are greater than the observed range of \HI\ gas surface densities in dwarf galaxies of $\ltsimeq10$ \msun~pc$^{-2}$ \citep{Swaters2002, Wong2002, Bigiel2008}. 
Using a higher SFE of 10\%, half of the inferred gas surface densities are still higher than the observed range of $\Sigma\rm_{HI}$ by factors of 9. 

The simplest explanation appears to be that a significant fraction of the gas is in molecular form but remains unobserved in most of these galaxies due to their low metallicities. Indeed, the elevated levels of recent SF are an unambiguous sign that our sample of galaxies have hosted significant reservoirs of cold molecular gas. 
In this low metallicity regime, the H$_2$ masses can be significant, even for very low CO luminosities. 
Thus, the lack of CO detections in our sample is not inconsistent with their having significant reservoirs of molecular gas. 
Coupled with limits on CO emission from the literature, the inferred molecular gas surface densities suggest X$\rm_{CO}$ factors of up to $40\times$ greater than Galactic measurements in these low metallicity and low mass galaxies.

\section{Acknowledgments}
Support for this work was provided by NASA through a ROSES grant (No.~NNX10AD57G). E.~D.~S. is grateful for partial support from the University of Minnesota. The authors thank the anonymous referee for helpful and constructive comments. K.~B.~W.~M. gratefully acknowledges Matthew, Cole, and Carling for their support. This research made use of NASA's Astrophysical Data System and the NASA/IPAC Extragalactic Database (NED) which is operated by the Jet Propulsion Laboratory, California Institute of Technology, under contract with the National Aeronautics and Space Administration. 

{\it Facilities:} \facility{Hubble Space Telescope}


\begin{deluxetable}{lccrcllcr}
\tabletypesize{\scriptsize}
\tablewidth{0pt}
\tablecaption{The Galaxy Sample and Properties\label{tab:galaxies}}
\tablecolumns{9}
\tablehead{
\colhead{}				&
\colhead{M$_B$}				&
\colhead{Distance}			&
\colhead{D$_{25}$ Area}			&
\colhead{Deproj. D$_{25}$ Area}		&
\colhead{Duration}			&
\colhead{$<$SFR$>$ over Burst}		&
\colhead{}				&
\colhead{}				\\
\colhead{Galaxy}			&
\colhead{(mag)}				&
\colhead{(Mpc)}				&
\colhead{in FOV (\%)}			&
\colhead{in FOV (kpc$^2$)}		&
\colhead{of Burst (Myr)}		&
\colhead{($\times10^{-3}$ \msun\ yr$^{-1}$)}&
\colhead{12$+$log(O/H)}			&
\colhead{Source}			\\
\colhead{(1)}				&	
\colhead{(2)}				&	
\colhead{(3)}				&	
\colhead{(4)}				&	
\colhead{(5)}				&	
\colhead{(6)}				&
\colhead{(7)}				&
\colhead{(8)}				&
\colhead{(9)}				
}

\startdata
Antlia		& -10.14 & 1.25	& 100\% & 0.47  & 636$\pm$185 	   & 0.18$\pm$0.02	& 7.39  & 1 \\
UGC 9128	& -12.45 & 2.24 & 100\% & 1.2   & 1300$\pm$300	   & 1.8 $\pm$0.01	& 7.74  & 2 \\
UGC 4483	& -12.68 & 3.2 	& 100\% & 1.0   & $>$810$\pm$190   & 2.5 $\pm$0.1	& 7.50  & 3 \\
NGC 4163	& -13.75 & 3.0 	& 100\% & 3.2   & 460$\pm$70	   & 4.5 $\pm$0.2	& 7.69  & 4 \\
UGC 6456 	& -13.85 & 4.3	& 100\% & 3.0   & $>$570$\pm$60	   & 3.5 $\pm$0.2	& 7.64  & 5 \\
NGC 6789 	& -14.60 & 3.6 	& 100\% & 1.7   & 480$\pm$70	   & 4.5 $\pm$0.1	& 7.77  & 4 \\
NGC 4068 	& -14.96 & 4.3	& 100\% & 19.5  & $>$459$\pm$50	   & 20. $\pm$1		& 7.84  & 4 \\
DDO 165 	& -15.19 & 4.6	& 100\% & 18.   & $>$1300$\pm$300  & 21. $\pm$1		& 7.80  & 4 \\
IC 4662 	& -15.39 & 2.4	& 100\% & 2.2   & $>$450$\pm$50	   & 28. $\pm$1		& 8.17  & 6 \\
ESO 154-023 	& -16.21 & 5.76	& 63\%  & 120   & $>$450$\pm$50	   & 53. $\pm$1		& 8.01  & 4 \\
NGC 2366	& -16.33 & 3.2	& 96\%  & 40.   & $>$450$\pm$50	   & 56. $\pm$1		& 8.19  & 7 \\
NGC 625 	& -16.26 & 3.9	& 63\%  & 35.   & 450$\pm$50	   & 18. $\pm$1		& 8.10  & 8 \\
NGC 784 	& -16.78 & 5.19	& 73\%  & 64.   & $>$450$\pm$50	   & 63. $\pm$2		& 8.05  & 4 \\
Ho II		& -16.92 & 3.4	& 44\%  & 21.   & 566$\pm$65	   & 51. $\pm$1		& 7.92  & 6 \\
NGC 5253 	& -16.98 & 3.5	& 100\% & 28.   & $>$450$\pm$50	   & 130 $\pm$2		& 8.10  & 9\\
NGC 6822	& -17.86 & 0.5	& 10\%  & 0.42  & $>$20$\pm$	   & 4.5 $\pm$0.4	& 8.11  & 10\\
NGC 4214	& -17.02 & 2.7	& 13\%  & 5.4   & 810$\pm$190	   & 46. $\pm$1		& 8.38  & 11\\
NGC 1569 	& -17.76 & 3.36	& 87\%  & 3.0   & $>$450$\pm$50	   & 76 $\pm$2		& 8.19  & 12\\
NGC 4449 	& -18.02 & 4.2	& 87\%  & 40.   & $>$450$\pm$50	   & 270 $\pm$5		& 8.21  & 13\\

\enddata
\tablecomments{Column (1) Galaxy name. Column (2) Absolute B magnitude of the galaxy. Column (3) Tip of the Red Giant Branch distance to the galaxy. Column (4) Indicates the percentage of the area defined by the B-band 25 mag arcsec$^{-2}$ isophotal diameter in the observational field of view (F.O.V.) form \citet{Karachentsev2004}. Column (5) Measures the deprojected D$_{25}$ area contained in the F.O.V. in kpc$^2$. Column (6) The starburst durations \citep{McQuinn2010b}. The durations are lower limits in 12 systems as the galaxies show on-going, elevated rates of SF. Column (7) The average SFRs over the duration of the starbursts \citep{McQuinn2010a} scaled from a Salpeter IMF to a Kroupa IMF by dividing by a factor of 1.59.}

\tablerefs{(1) \citet{Piersimoni1999}; (2) \citet{vanZee1997}; (3) \citet{Skillman1994}; (4) L-Z relation; \citet{Zaritsky1994, Tremonti2004, Lee2006}; (5) \citet{Croxall2009}; (6) \citet{Hidalgo2001a}; (7) \citet{Roy1996}; (8) \citet{Skillman2003}; (9) \citet{Kobulnicky1997a}; (10) \citet{Hidalgo2001b}; (11) \citet{Kobulnicky1996}; (12) \citet{Kobulnicky1997b}; (13)\citet{Skillman1989}}

\end{deluxetable}

\begin{sidewaystable}
\begin{center}
\caption{The SFR, Stellar, Gas Surface Densities, and X$\rm_{CO}$ Conversion Factors \label{tab:densities}}
\begin{tabular}{lccccccc}
\hline \hline
 & Measured & Measured & Inferred log $\Sigma\rm_{gas}$	& N$\rm_{H_2}$ ($\times10^{20}$ cm$^{-2}$) 	& Estimated M$\rm_{H2}$	& I$\rm_{CO}$& X$\rm_{CO}$\\
						&
log $\Sigma\rm_{SFR}$				&
log $\Sigma\rm_{*~burst}$ 			&
(\msun\ pc$^{-2}$)				&
Assumes					&
at onset of burst				&
(K km s$^{-1}$)				&
($\times10^{20}$)				\\
Galaxy					&
(\msun\ yr$^{-1}$ kpc$^{-2}$)			&
(\msun\ pc$^{-2}$)				&
SFE$=$10\%; He factor=1.33			&
$\Sigma\rm_{HI}=10$ \msun\ pc$^{-2}$		&
($\times10^6$ \msun)				&
from lit.					&
(cm$^{-2}$/(K km s$^{-1}$)			\\	
(1)                           &       
(2)                           &       
(3)                           &       
(4)                           &       
(5)                           &       
(6)                           &
(7)                           &
(8)\\

\hline
Antlia		& -3.41	$\pm$ 0.088	& -0.5	$\pm$0.05	& 0.37	&\nodata&\nodata& \nodata   & \nodata  \\
UGC 9128	& -2.82	$\pm$ 0.011	&  0.4	$\pm$0.03	& 1.24 	& $5$	&  9  	& \nodata   & \nodata \\
UGC 4483	& -2.61	$\pm$ 0.009	&  0.4	$\pm$0.06	& 1.26 	& $5$	&  8  	& $<0.195$  & $>26$ \\
NGC 4163	& -2.85 $\pm$ 0.005	& -0.1	$\pm$0.07	& 0.77	&\nodata&\nodata& \nodata   & \nodata \\
UGC 6456 	& -2.94	$\pm$ 0.005	& -0.1	$\pm$0.07	& 0.74	&\nodata&\nodata& \nodata   & \nodata \\
NGC 6789 	& -2.58	$\pm$ 0.005	&  0.2	$\pm$0.08	& 1.03 	& $4$	&  1   	& \nodata   & \nodata \\
NGC 4068 	& -3.00	$\pm$ 0.001	& -0.3	$\pm$0.02	& 0.57	&\nodata&\nodata& \nodata   & \nodata \\
DDO 165 	& -2.94	$\pm$ 0.001	&  0.3	$\pm$0.03	& 1.13 	& $2$	&  65  	& \nodata   & \nodata \\
IC 4662 	& -1.90	$\pm$ 0.003	&  0.8	$\pm$0.04	& 1.68 	& $23$	&  82 	& \nodata   & \nodata \\
ESO 154-023 	& -3.36	$\pm$ 0.001	& -0.7	$\pm$0.02 	& 0.22	&\nodata&\nodata& \nodata   & \nodata  \\
NGC 2366 	& -2.85	$\pm$ 0.001	& -0.2	$\pm$0.01	& 0.71	&\nodata&\nodata& $<0.6$    & \nodata \\
NGC 625 	& -3.28 $\pm$ 0.001	& -0.6	$\pm$0.05	& 0.27	&\nodata&\nodata& \nodata   & \nodata \\
NGC 784 	& -3.01 $\pm$ 0.001 	& -0.3	$\pm$0.02	& 0.58	&\nodata&\nodata& \nodata   & \nodata \\
Ho II		& -2.62	$\pm$ 0.001	&  0.2	$\pm$0.02	& 1.08 	& $1$	& 45  	& $<0.46$   & $>$3 \\
NGC 5253 	& -2.35 $\pm$ 0.001	&  0.4	$\pm$0.03	& 1.24 	& $5$	& 210  	& $0.725$   & 6 \\
NGC 6822	& -1.97	$\pm$ 0.1	& -0.4	$\pm$0.08	& 0.50	&\nodata&\nodata& $0.89$    & \nodata\\
NGC 4214	& -2.08	$\pm$ 0.002	&  0.9	$\pm$0.06	& 1.80 	& $33$	& 280   & 0.542	    & 61\\
NGC 1569 	& -1.60 $\pm$ 0.004	&  1.1	$\pm$0.04	& 1.96 	& $51$	& 240  	& 0.685	    & 74\\
NGC 4449 	& -2.17 $\pm$ 0.001	&  0.5	$\pm$0.05	& 1.39 	& $9$	& 590  	& 0.78	    & 12\\
\hline

\end{tabular}
\end{center}
\scriptsize{Col. (1) Galaxy name. Col. (2) The SFR surface density is equal to the average SFR over the life of the burst \citep{McQuinn2010b}, assuming a Kroupa IMF, and divided by the deprojected area out to the B-band 25 mag arcsec$^{-2}$ isophotal diameter or the field of view, whichever is smaller (see Table~\ref{tab:galaxies} for values). Col. (3) The stellar mass surface density is equal to the stellar mass formed during the burst divided by the the same area used for the $\Sigma\rm_{SFR}$ and converted to pc$^{-2}$. The stellar mass was calculated by integrating the SFRs over the duration of each starburst \citep{McQuinn2010b}. Col. (4) The inferred hydrogen gas surface density present at the onset of the burst is equal to the $\Sigma\rm_{*~burst}$ divided by an assumed SFE of $10\%$ and a factor of 1.33 to account for the abundance of Helium. Col. (5) The estimated number density of molecular hydrogen assuming an \HI\ gas surface density $=10$ \msun\ pc$^{-2}$, based on the gas surface densities in Col. (4). Col. (6) Molecular gas mass estimated at the onset of the starbursts in the areas listed in Table~\ref{tab:galaxies} based on the H$_2$ column densities in Col. (5). These values can be compared to the 10$^5-10^6$ \msun\ molecular gas mass typical of a Galactic GMC \citep{Blitz1993}. Col. (7) Intensity of CO emission including confirmed detections for NGC~5253, NGC~4214, NGC~1569 \citep{Taylor1998}, NGC~6822 \citep{Wilson1992}, and NGC~4449 \citep{Tacconi1985}, and observationally set upper limits for UGC~4483 \citep{Taylor1998}, NGC~2366 \citep{Hunter1993}, Ho~II \citep{Young1995}. Col. (8) X$\rm_{CO}$ factor estimated from the ratio of inferred H$_2$ column densities to measured CO luminosities for systems or observationally set upper limits, for the galaxies with inferred molecular gas content.}
\end{sidewaystable}

\end{document}